\documentclass[amsmath,amssymb,reprint,aip]{revtex4-2}
\usepackage{graphicx}
\usepackage{amsmath}
\usepackage{float}
\usepackage{hyperref}
\hypersetup{colorlinks=true, linkcolor=blue,
citecolor=blue, filecolor=blue, urlcolor=blue,}
\usepackage[caption=false]{subfig}
\usepackage{booktabs}

\begin{document}

\preprint{APS/123-QED}

\title{Hypergraph assortativity: a dynamical systems perspective}

\author{Nicholas W. Landry}
\email{nicholas.landry@colorado.edu}
\affiliation{Department of Applied Mathematics, University of Colorado at Boulder, Boulder, Colorado 80309, USA}
\author{Juan G. Restrepo}%
\email{juanga@colorado.edu}
\affiliation{Department of Applied Mathematics, University of Colorado at Boulder, Boulder, Colorado 80309, USA}

\date{\today}

\begin{abstract}
The largest eigenvalue of the matrix describing a network's contact structure is often important in predicting the behavior of dynamical processes. We extend this notion to hypergraphs and motivate the importance of an analogous eigenvalue, the expansion eigenvalue, for hypergraph dynamical processes. Using a mean-field approach, we derive an approximation to the expansion eigenvalue in terms of the degree sequence for uncorrelated hypergraphs. We introduce a generative model for hypergraphs that includes degree assortativity, and use a perturbation approach to derive an approximation to the expansion eigenvalue for assortative hypergraphs. We define the dynamical assortativity, a dynamically sensible definition of assortativity for uniform hypergraphs, and describe how reducing the dynamical assortativity of hypergraphs through preferential rewiring can extinguish epidemics. We validate our results with both synthetic and empirical datasets.
\end{abstract}

\maketitle
\begin{quotation}
Many models of epidemic spreading assume that individuals connect solely through pairwise interactions, which is often not the case. We define a model of epidemic spread that is mediated by group interactions, also known as higher-order interactions. Degree assortativity is the tendency of individuals with a similar number of connections to connect with each other more often than would be expected at random. When groups of individuals are connected in a degree assortative manner, epidemics are more likely to occur. We present a definition of degree assortativity for higher-order interaction networks that is meaningful for epidemic dynamics and show that epidemics can be extinguished by changing the assortative structure of these higher-order interaction networks.
\end{quotation}

\section{Introduction}

Complex social systems often exhibit assortative mixing \cite{newman_assortative_2002,newman_mixing_2003-1}, where individuals with similar characteristics connect with each other more often than it would be expected if they were connected at random. Assortativity has been extensively studied in network science \cite{boccaletti_complex_2006} and found to have significant effects on synchronization \cite{restrepo_mean-field_2014}, epidemic dynamics \cite{boguna_epidemic_2002,moreno_epidemic_2003}, stability \cite{brede_assortative_2005}, evolutionary game dynamics \cite{rong_roles_2007}, and general diffusion processes \cite{dagostino_robustness_2012}.

Recently, there has been much work on using hypergraphs to describe systems with interactions involving multiple agents \cite{battiston_networks_2020,benson_higher-order_2021,majhi_dynamics_2022}. Hypergraphs are useful to describe multi-way interactions in biology \cite{grilli_higher-order_2017}, social contagion \cite{de_arruda_social_2020,landry_effect_2020,iacopini_simplicial_2019}, synchronization \cite{skardal_abrupt_2019}, opinion models, infectious disease spread \cite{st-onge_social_2021}, and real data \cite{chodrow_configuration_2020}. Recently the pairwise notion of assortativity has been extended to hypergraphs for categorical node labels \cite{kaminski_clustering_2019,amburg_clustering_2020,chodrow_annotated_2020} and continuous attributes \cite{chodrow_configuration_2020}. Assortativity on hypergraphs can provide different insights on the structure of the interactions than assortativity on the pairwise network projection \cite{chodrow_configuration_2020} and, as we will show, affect the outcome of hypergraph dynamical processes.

A fundamental problem when studying dynamics on networks is to determine how structural characteristics of the network affect the dynamical behavior. Many dynamical properties such as the onset of epidemic spreading \cite{wang_epidemic_2003}, synchronization \cite{restrepo_onset_2005}, and percolation \cite{restrepo_weighted_2008} are determined by the largest eigenvalue of the network's adjacency matrix (or, in some cases, of the non-backtracking matrix \cite{karrer_percolation_2014}). In turn, this eigenvalue is affected by the network's degree distribution and assortative mixing properties \cite{restrepo_approximating_2007} as well as other structural characteristics. In this paper we show how the expansion eigenvalue, a suitably generalized eigenvalue for hypergraphs, is similarly modified by the assortative properties of the hypergraph. This eigenvalue has been shown to determine the extinction threshold for the susceptible--infected--susceptible (SIS) model on hypergraphs \cite{higham_epidemics_2021}, and we believe it will also prove useful in relating hypergraph assortative mixing patterns to other dynamical processes.

Our approach is as follows: first, we define and motivate the importance of the expansion eigenvalue on dynamical processes; second, we derive a mean-field approximation of this eigenvalue for hypergraphs without assortativity; third, we present a generative model for assortative hypergraphs; fourth, we employ a perturbation approach to derive the effect of degree-degree mixing on the eigenvalue and define the dynamical assortativity; and lastly, we show how our results can be used to modify hypergraph dynamics through preferential rewiring of hyperedges.

\section{Preliminaries}

We start by defining terminology. A hypergraph is a mathematical object that describes group interactions among a set of nodes. We represent it as $H=(V,E)$, where $V$ is the set of nodes and $E$ is the set of hyperedges, which are subsets of $V$ and represent unordered interactions of arbitrary size. We call a hyperedge with cardinality $m$ an $m$-hyperedge and a hypergraph with only $m$-hyperedges an m-uniform hypergraph. It is useful to consider weighted hypergraphs, where each hyperedge $e$ has an associated positive weight $\beta_e$. We define the hyperdegree sequence as in Ref.~\cite{landry_effect_2020}, where the $m$th order hyperdegree of node $i$, $k^{(m)}_i$, is the number of $m$-hyperedges to which it belongs.

We now define the expansion eigenvalue and discuss its relevance to dynamical processes on hypergraphs. For a weighted hypergraph, the expansion eigenvalue $\lambda$ and associated eigenvector ${\bf u}$ are defined by the eigenvalue equation
\begin{equation}
\lambda u_i = \sum_{e=\{i,i_1,\dots,i_{m-1}\}\in E} \beta_e(u_{i_1} + \dots + u_{i_{m-1}}),
\label{eqn:expansion_eigenvalue}
\end{equation}
where $\lambda$ and ${\bf u}$ are the Perron-Frobenius eigenvalue and eigenvector of the nonnegative matrix associated to linear equation~\eqref{eqn:expansion_eigenvalue}.

\subsection{Motivation}

Here we present some applications of the expansion eigenvalue. First, just like the Perron-Frobenius eigenvector of a network adjacency matrix represents eigenvector centrality \cite{bonacich_power_1987}, in the unweighted case (i.e., $\beta_e=1$ for every hyperedge $e$), the eigenvector ${\bf u}$ corresponds to the Clique motif Eigenvector Centrality, a generalization of eigenvector centrality for hypergraphs \cite{benson_three_2019}. Second, just as the largest eigenvalue of a network's adjacency matrix influences network dynamics, the expansion eigenvalue plays an important role in dynamical processes on hypergraphs. For example, consider an SIS process on a hypergraph, where a healthy node can get infected via a hyperedge $e$ to which it belongs at rate $\beta_e$ if at least one other node in $e$ is infected (the case referred to as individual contagion in Ref.~\cite{landry_effect_2020}) and heals spontaneously at rate $\gamma$. As discussed in Ref.~\cite{higham_epidemics_2021} in Theorem 9.1, the extinction threshold for the exact stochastic process can be bounded above by that for the mean-field dynamics. The mean-field equation for $x_i$, the probability that node $i$ is infected, is given by
\begin{align}
\frac{dx_i}{dt}&=-\gamma x_i + (1-x_i)\sum_{e=\{i,i_1, \dots, i_{m-1}\}\in E} \beta_e\nonumber&\\
&\times [1 - (1 - x_{i_1})\dots (1 - x_{i_{m-1}})].
\end{align}

By inspection, $x_i = 0$ for all $i$ is always a fixed point of this equation. We write an ODE for linear perturbations around this equilibrium to derive conditions for the system's stability. To first order, the equation for the perturbations, $\delta x_i$, is
\begin{equation}
\frac{d (\delta x_i)}{dt}=-\gamma (\delta x_i) + \sum_{\{i,i_1, \dots, i_{m-1}\}\in E} \beta_e (\delta x_{i_1} + \dots + \delta x_{i_{m-1}}),
\label{eq:perturbation_ode}
\end{equation}

If we assume $\delta x_i = u_i e^{r t}$, then
\begin{equation}
(r +\gamma) u_i = \lambda u_i = \sum_{\{i,i_1, \dots, i_{m-1}\}\in E} \beta_e (u_{i_1} + \dots + u_{i_{m-1}}),
\end{equation}
where $\lambda$ is the expansion eigenvalue and so, $r = \lambda -\gamma$. Therefore, a sufficient condition for epidemic extinction is $\gamma > \lambda$ \cite{higham_epidemics_2021}. For an $m$-uniform hypergraph with $\beta_e=\beta_m$, the extinction threshold is $\beta_m/\gamma < 1/\lambda$, where $\lambda$ is the expansion eigenvalue of the unweighted hypergraph.

If we rewrite the last term of Eq.~\eqref{eq:perturbation_ode} as a sum over uniform hypergraphs, then
\begin{equation}
\frac{d(\delta x_i)}{dt} = -\gamma (\delta x_i) + \sum_{m=2}^M  (m-1)\left(W^{(m)}\boldsymbol{\delta x}\right)_i,
\end{equation}
where $W^{(m)}$ is the weighted version of the clique motif matrix defined in Ref.~\cite{benson_three_2019} and $\boldsymbol{\delta x} = [\delta x_1,\dots, \delta x_N]$. We can define $W = \sum_{m=2}^M(m-1) W^{(m)}$ as a linear operator with eigenvalue $\lambda$ and as before, the extinction threshold is $\gamma > \lambda$. In addition, we can determine the relative importance of a node $i$ with respect to this contagion model (in terms of its probability of infection at the onset of the epidemic) from the $i$th entry of the associated eigenvector.

Finally, the importance of the expansion eigenvalue in spreading processes can be understood from the fact that in the unweighted case the number of nodes reachable via hyperedges from a given starting node in $\ell$ steps grows asymptotically as $\lambda^{\ell}$ as demonstrated in Ref.~\cite{benson_three_2019}.

\subsection{Hypergraph model}

In this paper we will consider random hypergraphs that are constructed from a prescribed hyperdegree sequence $\{{\bf k}_1,\dots,{\bf k}_N\}$, where $N$ is the number of nodes, ${\bf k}_i=[k^{(2)}_i,\dots,k^{(M)}]$ is the target hyperdegree of node $i$, and $M$ is the maximum hyperedge size. The hypergraph is then constructed by creating a hyperedge $\{i_1,\dots,i_m\}$ with probability $f_m({\bf k}_{i_1},\dots,{\bf k}_{i_m})$, where the functions $f_m$ specify the assortative mixing properties of the hypergraph.

This generative model can produce hypergraphs with heterogeneity in the node hyperdegrees and correlations between hyperdegrees of connected nodes. It can also be easily generalized to account for assortativity by additional nodal variables such as community labels or dynamical parameters. Its main limitation is that it does not capture connection patterns that are determined by structures beyond a node's immediate connections (e.g., the model cannot account for hyperedges of size 3 that occur only when there is a clique of 3 nodes connected by links, as one would see in a simplicial complex). Nevertheless, this generative model is a versatile and tractable null model to explore the effect of hypergraph structure on various hypergraph metrics, in particular the expansion eigenvalue.

\section{Mean-Field Approach}

\subsection{Uncorrelated m-uniform case}

We start by deriving a mean-field approximation for the expansion eigenvalue $\lambda$ in the case where nodes are connected with hyperedges completely at random (as in the hypergraph configuration model \cite{young_construction_2017, courtney_generalized_2016,  landry_effect_2020,chodrow_configuration_2020}), which we call the {\it uncorrelated case}, before considering hypergraphs with degree assortativity. In the uncorrelated case, the function $f_m$ is given by $f_m({\bf k}_{i_1},\dots,{\bf k}_{i_m})=f^{(0)}_m(k^{(m)}_{i_1},\dots,k^{(m)}_{i_m}) = (m-1)!k^{(m)}_{i_1}\dots k^{(m)}_{i_m}/(N\langle k^{(m)}\rangle)^{m-1}$, where we define $\langle x^p \rangle = \sum_{i=1}^N x_i^p/N$, corresponds to the case where nodes are connected with hyperedges completely at random if the hyperdegree of node $i$ is ${\bf k}_i$. For simplicity, from now on we will consider an unweighted $m$-uniform hypergraph, and will denote $k^{(m)}_i$ by $k_i$ and refer to it as the degree of node $i$. Now we assume that all nodes with the same degree are statistically equivalent and that the eigenvector entry of node $i$ depends only on its degree, i.e., $u_i \to u_{k_i}$. In Section~\ref{sec:discussion}, we discuss the limitations of this approach. Henceforth $\lambda$ will denote the mean-field approximation to the expansion eigenvalue for convenience unless explicitly stated otherwise. Defining $N(k)$ to be the number of nodes with degree $k$ such that $P(k) = N(k)/N$ is the degree distribution, the equation defining the expansion eigenvalue can be written as
\begin{align}
\lambda u_k&=\frac{1}{(m-1)!}\sum_{k_1, \dots, k_{m-1}} N(k_1)\dots N(k_{m-1}) \nonumber \\
&\times f^{(0)}_m(k, k_1, \dots, k_{m-1})(u_{k_1} + \dots + u_{k_{m-1}}).
\label{eqn:eig_eqn_config_model}
\end{align}

By symmetry of the function $f_m^{(0)}$, we get
\begin{equation}
\lambda u_k = \left[ (m-1) \sum_{k_1}P(k_1) \frac{k_1 \, u_{k_1}}{\langle k\rangle}\right] k,
\label{eqn:eig_eqn_config_simplified}
\end{equation}
and multiplying both sides by $k \, P(k)/\langle k\rangle$ and summing over $k$, we obtain for the uncorrelated case
\begin{equation}
\lambda = (m-1)\frac{\langle k^2\rangle}{\langle k\rangle},
\end{equation}
and $u_k\propto k$ from Eq.~\eqref{eqn:eig_eqn_config_simplified}.

\subsection{Derivation of the non-uniform uncorrelated expansion eigenvalue}

We now relax the assumption of an $m$-uniform hypergraph and consider an uncorrelated hypergraph with hyperedges of sizes $m = 2,\dots ,M$ and hyperedge weights of the form $\beta_e=\beta_{|e|}$. The expansion eigenvalue equation can be written as
\begin{equation}
\lambda u_i = \sum_{m=2}^M \beta_m \sum_{\{i,i_1,\dots,i_{m-1}\}\in E}(u_{i_1} + \dots + u_{i_{m-1}}).
\label{eqn:expansion_eigenvalue2}
\end{equation}
The degree-based mean-field eigenvalue equation, where we assume $u_i=u_{{\bf k}_i}$, can be written as
\begin{align}
\lambda u_{\bf k}&=\sum_{m=2}^M \beta_m \frac{1}{(m-1)!}\sum_{{\bf k}_1, \dots, {\bf k}_{m-1}} N({\bf k}_1)\dots N({\bf k}_{m-1})\nonumber\\
&\times f_m({\bf k},{\bf k}_{1},\dots,{\bf k}_{m-1})(u_{{\bf k}_1} + \dots + u_{{\bf k}_{m-1}}).
\end{align}
Focusing on the uncorrelated case, we assume that 
\begin{align*}
f_m({\bf k}, {\bf k}_1, \dots, {\bf k}_{m-1})&=f^{(0)}_m(k^{(m)}, k^{(m)}_1, \dots, k^{(m)}_{m-1})\\
&= \frac{(m-1)! k^{(m)} k_1^{(m)} \dots k^{(m)}_{m-1}}{(N\langle k^{(m)}\rangle)^{m-1}},
\end{align*}
so
\begin{align*}
\lambda u_{\bf k}&=\sum_{m=2}^M \beta_m \sum_{{\bf k}_1, \dots, {\bf k}_{m-1}} N({\bf k}_1)\dots N({\bf k}_{m-1})\nonumber\\
&\times \frac{k^{(m)} k^{(m)}_1\dots k^{(m)}_{m-1}}{(N\langle k^{(m)}\rangle)^{m-1}}(u_{{\bf k}_1} + \dots + u_{{\bf k}_{m-1}}),
\end{align*}
and from symmetry,
\begin{equation}
\lambda u_{\bf k}=\sum_{m=2}^M k^{(m)} \beta_m (m-1) \sum_{{\bf k}_1}P({\bf k}_1) \frac{k^{(m)}_1 u_{{\bf k}_1}}{\langle k^{(m)}\rangle}.
\label{eq:non-uniform_mf_simplified}
\end{equation}
From Eq.~\eqref{eq:non-uniform_mf_simplified}, we can see that $u_{{\bf k}}$ must be a linear combination of $k^{(m)}$. We assume an ansatz of the form
\begin{equation}
u_{\bf k} = \sum_{m=2}^M v_m k^{(m)} ={\bf k}^T{\bf v}
\end{equation}
where ${\bf v} = (v_2, \dots, v_M)$ is an unknown vector of nonnegative weights. Renaming the summation indices and evaluating this ansatz in the eigenvalue equation,
\begin{align*}
\lambda \sum_{j=2}^M v_j k^{(j)} &= \sum_{i=2}^M k^{(i)} \beta_i (i-1)\\
&\times\sum_{{\bf k}_1} P({\bf k}_1) \frac{k^{(i)}_1 \sum_{j=2}^M v_j k_1^{(j)}}{\langle k^{(i)}\rangle}.
\end{align*}
Changing the order of summation,
\begin{align*}
\lambda{\bf k}^T{\bf v}&=\sum_{i=2}^M \sum_{j=2}^M k^{(i)} \frac{\beta_i (i-1)}{\langle k^{(i)}\rangle} v_j\sum_{{\bf k}_1} P({\bf k}_1) k^{(i)}_1 k_1^{(j)},\\
&=\sum_{i=2}^M \sum_{j=2}^M k^{(i)} \frac{\beta_i (i-1)\langle k^{(i)} k^{(j)}\rangle}{\langle k^{(i)}\rangle} v_j,\\
&={\bf k}^T K {\bf v}.\\
\end{align*}
We call $K$ the degree-size correlation matrix, with entries $K_{ij} = \beta_i (i-1)\langle k^{(i)} k^{(j)}\rangle/\langle k^{(i)}\rangle$ which we call the inter-size correlations. In Ref.~\cite{sun_higher-order_2021}, the authors derive a similar matrix for higher-order percolation processes. Generically (when ${\bf k}$ is not orthogonal to the range of $K-\lambda I$), this equation has a solution if and only if $\lambda$ and ${\bf v}$ solve the eigenvalue equation $\lambda {\bf v} = K{\bf v}$. Notice that in the $m$-uniform case, we recover the expression we previously derived. Consider the network formed by specifying hyperedge sizes ($m=2,\dots,M$) to be the nodes and constructing a link between two sizes $m_1$ and $m_2$ if at least one node in the original hypergraph is a member of a hyperedge of size $m_1$ and a hyperedge of size $m_2$. $K$ is irreducible if and only if this network is strongly connected. If this is the case, by the Perron-Frobenius theorem the eigenvalue with largest magnitude is positive and has a corresponding positive eigenvector, and they correspond, respectively, to $\lambda$ and ${\bf v}$.

\subsection{Perturbation approach for the correlated case}

In contrast with the uncorrelated case, we now assume that nodes are connected with an arbitrary function $f_m$ determining the connection probability. We define
\begin{align}
f_m(k_1,\dots,k_m) &= f^{(0)}_m(k_1,\dots,k_m)\left[1 + \epsilon g_m(k_1, \dots, k_m)\right],
\label{eqn:fm}
\end{align}
where $\epsilon$ is a parameter which will later assume to be small and $g_m$ an assortativity function for $m$-uniform hypergraphs. The assortativity function $g_m(k_1,\dots,k_m)$ determines how likely it is that nodes with degrees $k_1,\dots, k_m$ are joined by a $m$-hyperedge; if $\epsilon g_m > 0$ ($\epsilon g_m < 0$) it is more (less) likely than it would be expected if they were connected at random. In order to preserve the expected degree sequence, $g_m$ must satisfy $\sum_{k_1,\dots,k_m} f_m^{(0)}(k_1,\dots, k_m) g_m(k_1,\dots, k_m) = 0$.

We now assume that the parameter $\epsilon$ is small and develop perturbative approximations to the eigenvalue $\lambda$ and its eigenvector $u_k$. To first order these approximations are
\begin{equation}
\begin{aligned}
\lambda &= \lambda^{(0)} + \epsilon \lambda^{(1)},\\
u_k &= u_k^{(0)} + \epsilon u_k^{(1)},
\end{aligned}
\label{eq:lambda_u_first_order}
\end{equation}
where $\lambda^{(0)}=(m-1)\langle k^2\rangle/\langle k\rangle$ and $u_k^{(0)} = \alpha k$, where $\alpha$ is an arbitrary constant.

Replacing $f^{(0)}_m$ on the right-hand side of Eq.~\eqref{eqn:eig_eqn_config_model} with the $f_m$ in Eq.~\eqref{eqn:fm}, using Eq.~\eqref{eq:lambda_u_first_order}, assuming symmetry of $f_m$, multiplying by $kP(k)/(N\langle k\rangle)$, summing over $k$, and canceling the zero-order terms, we obtain to first order (see Appendix \ref{sec:appendix_eigenvalue} for more detailed calculations)
\begin{align}
\lambda^{(1)} &= (m-1)\frac{\langle k\rangle}{\langle k^2\rangle}\sum_{k,k_1, \dots, k_{m-1}} N(k) N(k_1)\dots N(k_{m-1}) \nonumber\\
&\times \frac{k^2 \,k_1^2 \,k_2 \dots k_{m-1}}{(N\langle k\rangle)^m} g_m(k, k_1,\dots, k_{m-1}).
\end{align}

Removing the reference to $g_m$ using the relation in Eq.~\eqref{eqn:fm} we find
\begin{equation}
\epsilon\lambda^{(1)} = (m-1)\frac{\langle k\rangle\langle kk_1\rangle_{E}}{\langle k^2\rangle} - \lambda^{(0)},
\end{equation}
where $\langle k k_1\rangle_E$ is the mean pairwise product of degrees over all possible 2-node combinations in each hyperedge in the hypergraph, $\langle k k_1\rangle_E = \sum_{e\in E}\sum_{\{i,j\}\subseteq e} k_i k_j/\left(|E|\binom{m}{2}\right)$.

\begin{figure}
    \centering
    \includegraphics[width=8.6cm]{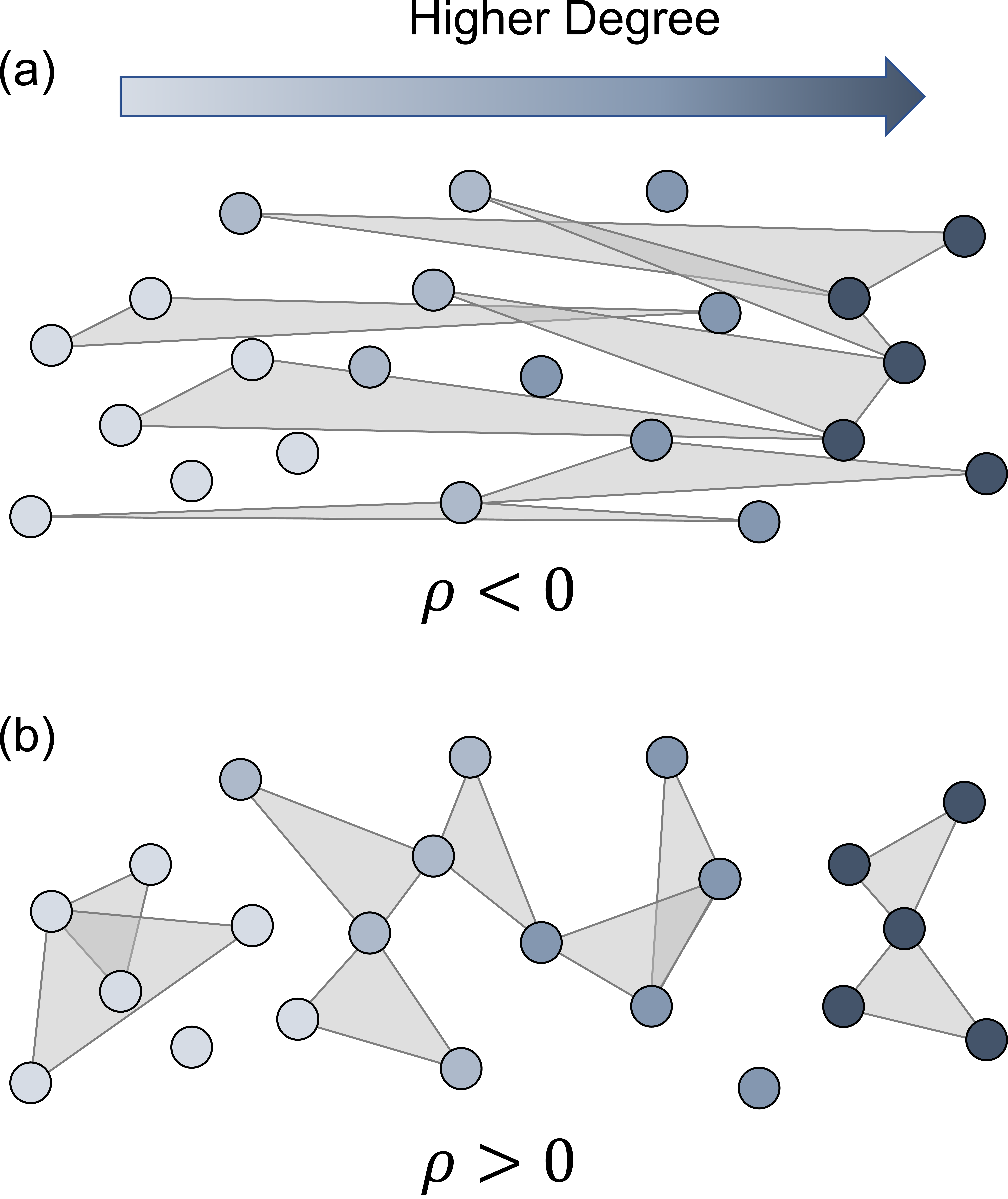}
    \caption{An illustration of disassortative and assortative 3-uniform hypergraphs. The color of the nodes indicates their degree, with low-degree nodes on the left and high-degree nodes on the right. For a given degree sequence, the term $\langle k k_1\rangle_E$ (the average pairwise product) determines $\rho$ and on average, (a) hyperedges containing nodes with dissimilar degrees decrease this term leading to disassortative hypergraphs and (b) hyperedges containing nodes of similar degree increase this term leading to assortative hypergraphs.}
    \label{fig:rho_illustration}
\end{figure}

Therefore, the expansion eigenvalue can be written, to first order, as
\begin{align}
\lambda &= \lambda^{(0)} + \epsilon \lambda^{(1)} = (m-1)\frac{\langle k\rangle\langle kk_1\rangle_{E}}{\langle k^2\rangle},\nonumber \\
&=\lambda^{(0)}(1+ \rho),
\end{align}
where we defined
\begin{equation}
\rho = \frac{\langle k\rangle^2\langle k k_1\rangle_E}{\langle k^2\rangle^2} - 1.
\label{eqn:rho}
\end{equation}
We refer to $\rho$ as the dynamical assortativity for its relation to hypergraph dynamics. One can verify that the expected value of $\rho$ for an uncorrelated hypergraph is 0. Interestingly, to first order the expansion eigenvalue does not depend on the particular assortativity function $g_m$ used, but only on the average of pairwise products of the degrees belonging to the same hyperedge. A schematic of disassortative ($\rho < 0$) and assortative ($\rho > 0$) hypergraphs is shown in Fig.~\ref{fig:rho_illustration}.

\section{Numerical Results}

\subsection{Approximating the eigenvalue}

We validate our results with numerical simulations on both synthetic and empirical hypergraphs. For both types of data, we modify the dynamical assortativity of the datasets by performing preferential double hyperedge swaps on the hypergraphs.

For each dataset hypergraph $H$,  we focus on an $m$-uniform partition $H_m$ (i.e., we only consider its hyperedges of size $m$). We set a target dynamical assortativity $\hat{\rho}$ and swap edges as follows. We choose two hyperedges $e_1 = \{i_1,i_2,\dots,i_m\}$ and $e_2 = \{j_1,j_2,\dots,j_m\}$ and a node from each uniformly at random, say $i_1$ and $j_1$. Then we consider the rewired hypergraph $H'_m$ obtained by replacing $e_1$ and $e_2$ with $e'_1 = \{j_1,i_2,\dots,i_m\}$ and $e'_2 = \{i_1,j_2,\dots,j_m\}$ respectively. If the assortativity of $H'_m$ with this hyperedge swap, $\rho'$, reduces the difference between the current assortativity, $\rho$, and the desired assortativity, $\hat{\rho}$, the swap is accepted and we set $H_m=H'_m$. To ensure that the algorithm explores the space of possible hypergraphs, we accept hyperedge swaps which increase the difference between the desired assortativity and the current assortativity with probability $e^{-[(\hat{\rho} - \rho)^2 - (\hat{\rho} - \rho')^2]/T}$ (we set $T=10^{-5}$). We terminate the algorithm when $\vert \rho - \hat{\rho}\vert$ is smaller than a prescribed tolerance or when a maximum number of hyperedge swaps have been performed (we used a tolerance of $10^{-2}$ and $10^6$ maximum hyperedge swaps).

For the synthetic hypergraph, we constructed a 3-uniform configuration model (CM) hypergraph of size $N=10^5$ according to the algorithm described in Ref.~\cite{landry_effect_2020} with a degree sequence drawn from a truncated power-law distribution, $P(k)\propto k^{-3}$ on $[10, 100]$. We also used the tags-ask-ubuntu (TAU), congress-bills (CB), and Eu-Emails (EE) hypergraph datasets from Refs.~\cite{benson_data_2021,fowler_connecting_2006,fowler_legislative_2006}, filtered to only include hyperedges of size 3. The characteristics of these datasets are described in Table~\ref{tab:datasets}.

\begin{table}[t]
\begin{ruledtabular}
\begin{tabular}{@{}lccc@{}}
\textbf{Dataset} & $N$    & $\langle k^{(3)} \rangle$ & $P(k^{(3)})$ \\
\hline\\
CM & $10^4$ & $18.2$ & \raisebox{-0.5\totalheight}{\includegraphics[width=5cm]{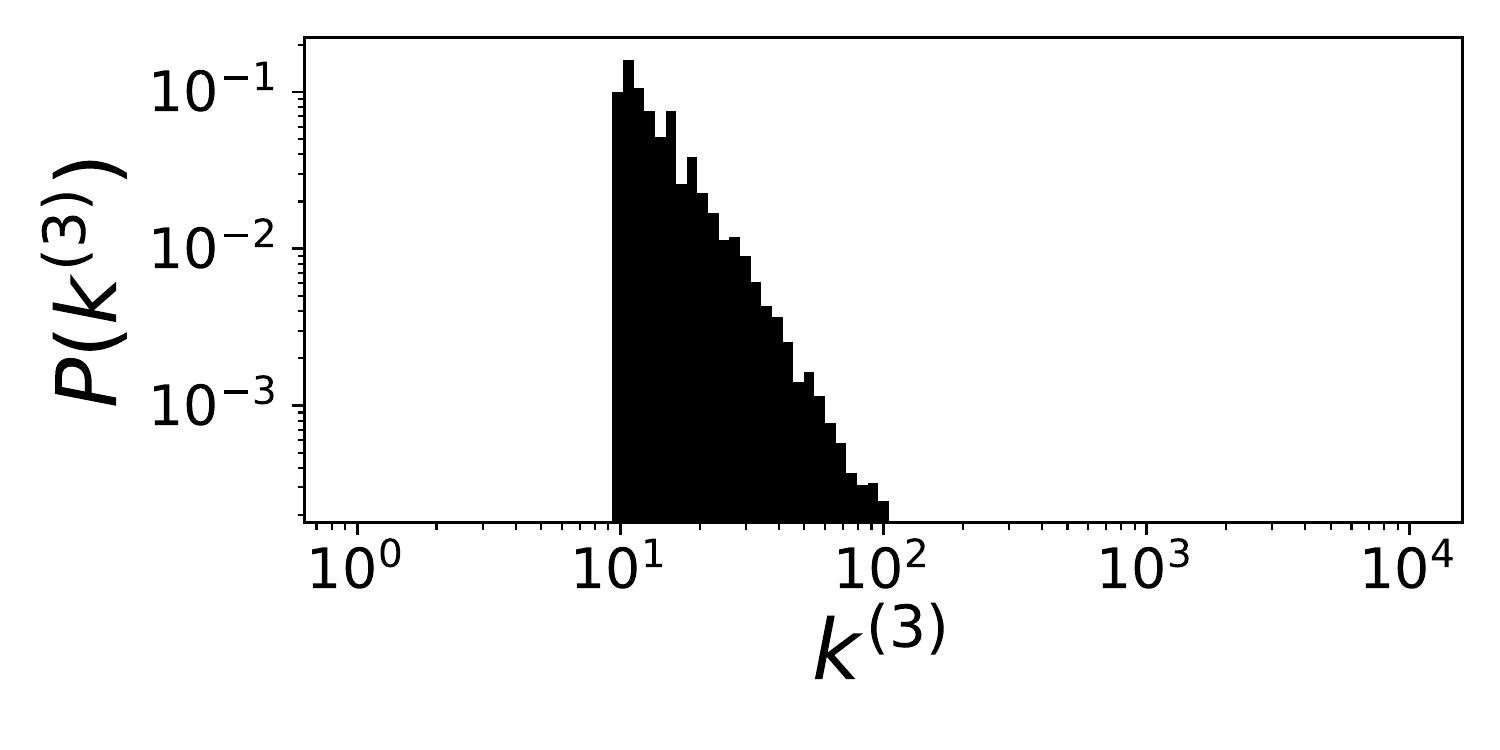}} \\
TAU & $3029$ & $71.2$ & \raisebox{-0.5\totalheight}{\includegraphics[width=5cm]{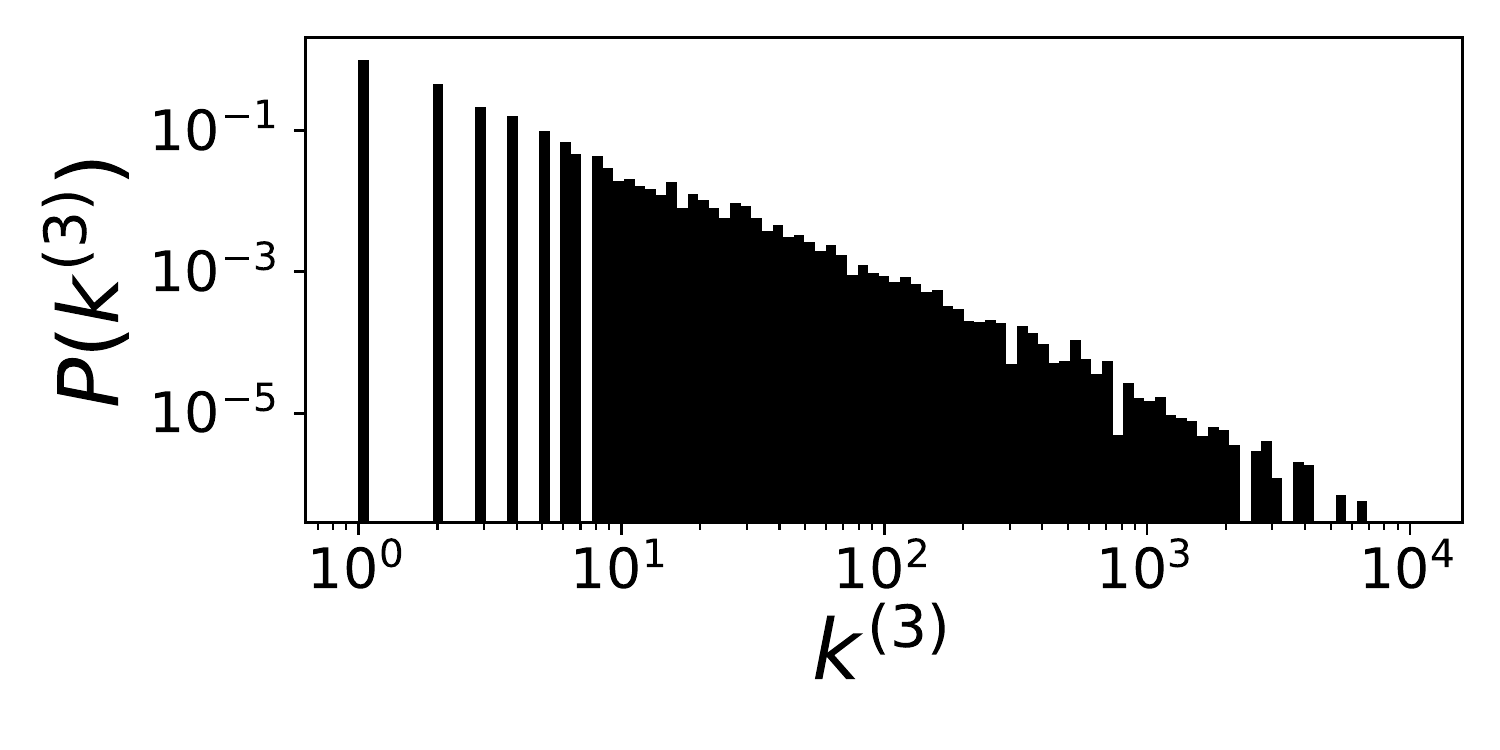}} \\
CB & $1718$ & $20.6$ & \raisebox{-0.5\totalheight}{\includegraphics[width=5cm]{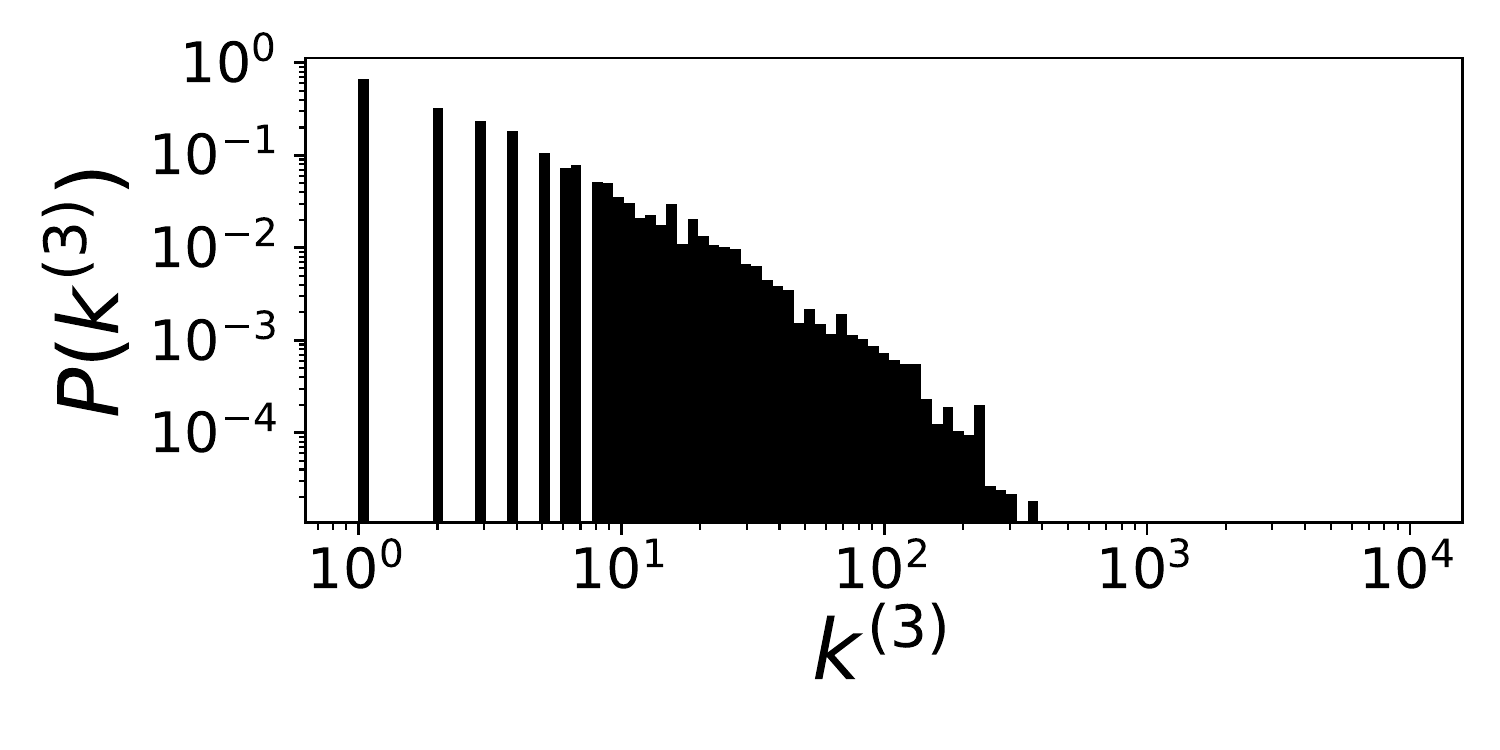}}\\
EE & $998$ & $53.7$ & \raisebox{-0.5\totalheight}{\includegraphics[width=5cm]{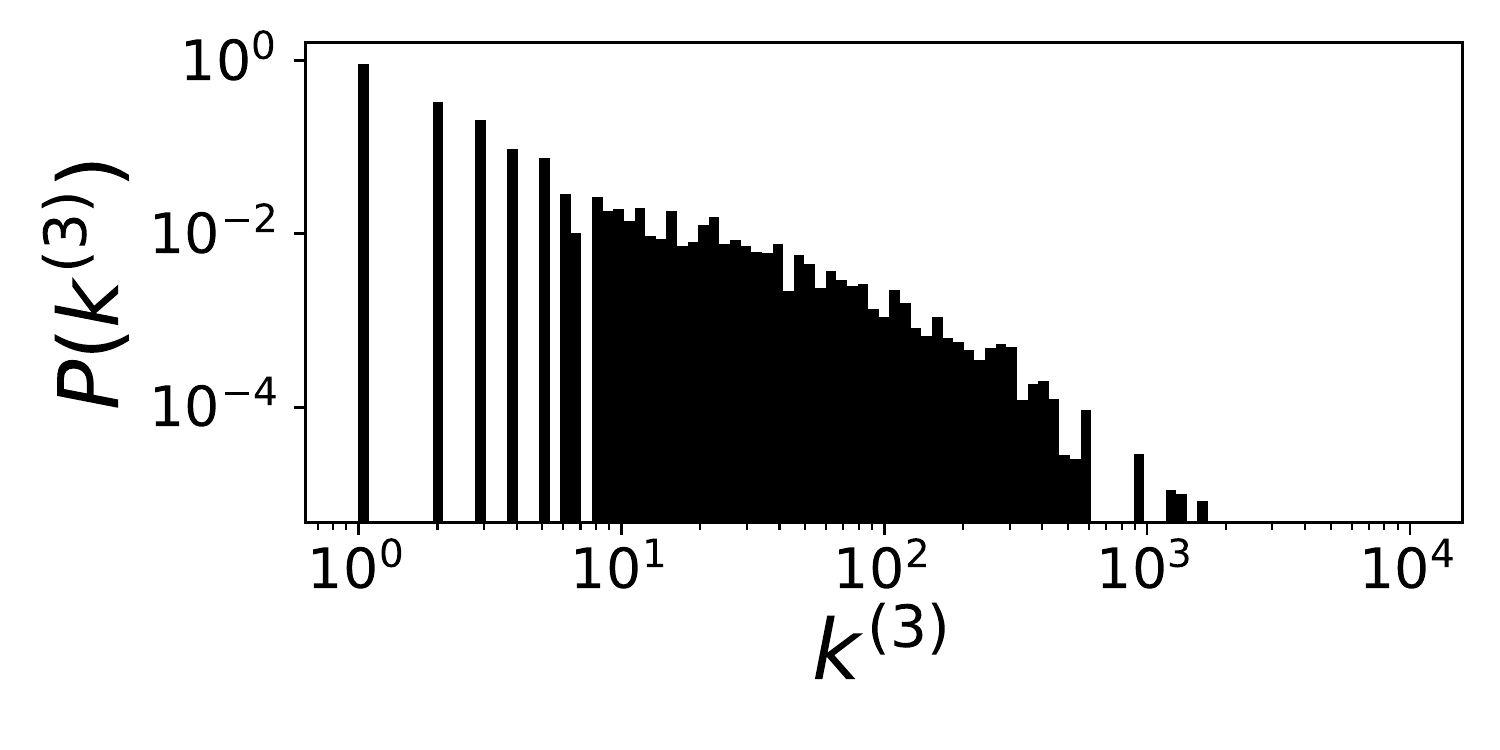}}\\ \bottomrule
\end{tabular}
\end{ruledtabular}
\caption{\label{tab:datasets}Characteristics of the 3-uniform hypergraph datasets used.}
\end{table}

\begin{figure*}
    \centering
    \includegraphics[width=17.2cm]{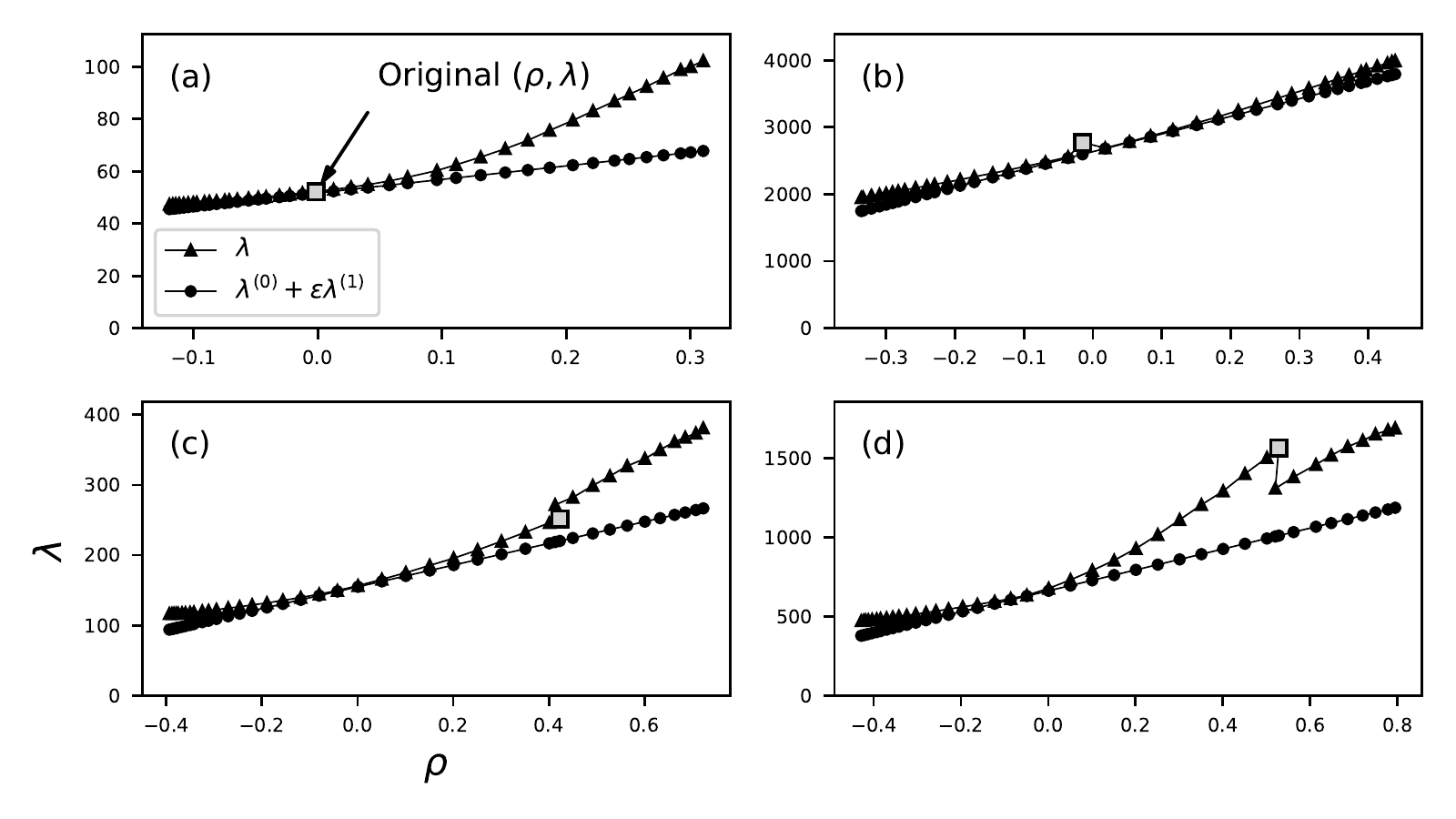}
    \caption{A comparison of the actual expansion eigenvalue $\lambda$ (connected triangles) to the first-order approximation of the eigenvalue $\lambda^{(0)} +\epsilon \lambda^{(1)}$ (connected circles) for (a) the configuration model, (b) the tags-ask-ubuntu dataset, (c) the congress-bills dataset, and (d) the Eu-Emails dataset. The square marker denotes the original $(\rho,\lambda)$ value of the dataset. Details of the characteristics of these datasets can be found in Table~\ref{tab:datasets}.}
    \label{fig:eigenvalue_vs_rho}
\end{figure*}

In Fig.~\ref{fig:eigenvalue_vs_rho}, the expansion eigenvalue $\lambda$ calculated numerically via the power method from Eq.~\eqref{eqn:expansion_eigenvalue} (connected triangles) and the first-order approximation $\lambda^{(0)}+\epsilon\lambda^{(1)}$ (connected circles) are plotted as a function of $\rho$ for the four datasets mentioned above. For each dataset, the starting point [i.e., the point $(\rho,\lambda)$ for the original hypergraph] is shown with a square marker. For the synthetic hypergraph (a), as expected, the first order approximation works well for small values of dynamical assortativity. For the TAU dataset (b) the agreement is even better than for the synthetic dataset for larger values of $\rho$. Interestingly, for the CB (c) and EE (d) datasets, and to a much lesser extent for the TAU dataset, the value of $\lambda$ changes sharply when first increasing (CB dataset and EE datasets), or both increasing and decreasing (TAU dataset) the assortativity. We hypothesize that initial hyperedge swaps might be destroying other structure (such as community structure, clustering, or assortative mixing by unaccounted attributes), causing $\lambda$ to change abruptly as this structure is destroyed, and then to change slowly as the effects of changing the assortativity dominate. We note that there appear to be limitations to the extent to which $\rho$ can be modified. This is similar to the limitations to the values of assortativity that networks and hypergraphs can achieve \cite{veldt_higher-order_2021,litvak_uncovering_2013,hofstad_degree-degree_2014,cinelli_network_2020}.

In all cases, we see that rewiring the hypergraph to increase the average value of $\langle k k_1\rangle_E$ (or, equivalently, $\rho$) has a dramatic effect on the expansion eigenvalue. For example, for the EE dataset $\lambda$ can be reduced threefold by the rewiring process. Thus, hypergraph rewiring might be a useful theoretical tool to control dynamical processes that depend on the expansion eigenvalue.

\subsection{Extinguishing epidemics}

Lastly, we show how modifying the dynamical assortativity by rewiring hypergraphs can extinguish an epidemic. As an example, consider a hypergraph SIS contagion spreading amongst groups of size $m$ at a \textit{fixed} rate $\beta_m$. In Ref.~\cite{higham_epidemics_2021}, the authors derive a sufficient condition for epidemic extinction for such models. For $m$-uniform hypergraphs and $\beta_e = \beta_m$, the extinction threshold for the individual contagion model is $\beta_m < \beta_m^c = \gamma/\lambda$. By decreasing $\lambda$ through hyperedge swaps and thus increasing $\beta_m^c$ so that $\beta_m^c >\beta_m$, the epidemic can be extinguished. (Note, however, that this is a sufficient condition; $\beta_m > \beta_m^c$ may not lead to an epidemic.)

\begin{figure}
    \centering
    \includegraphics[width=8.6cm]{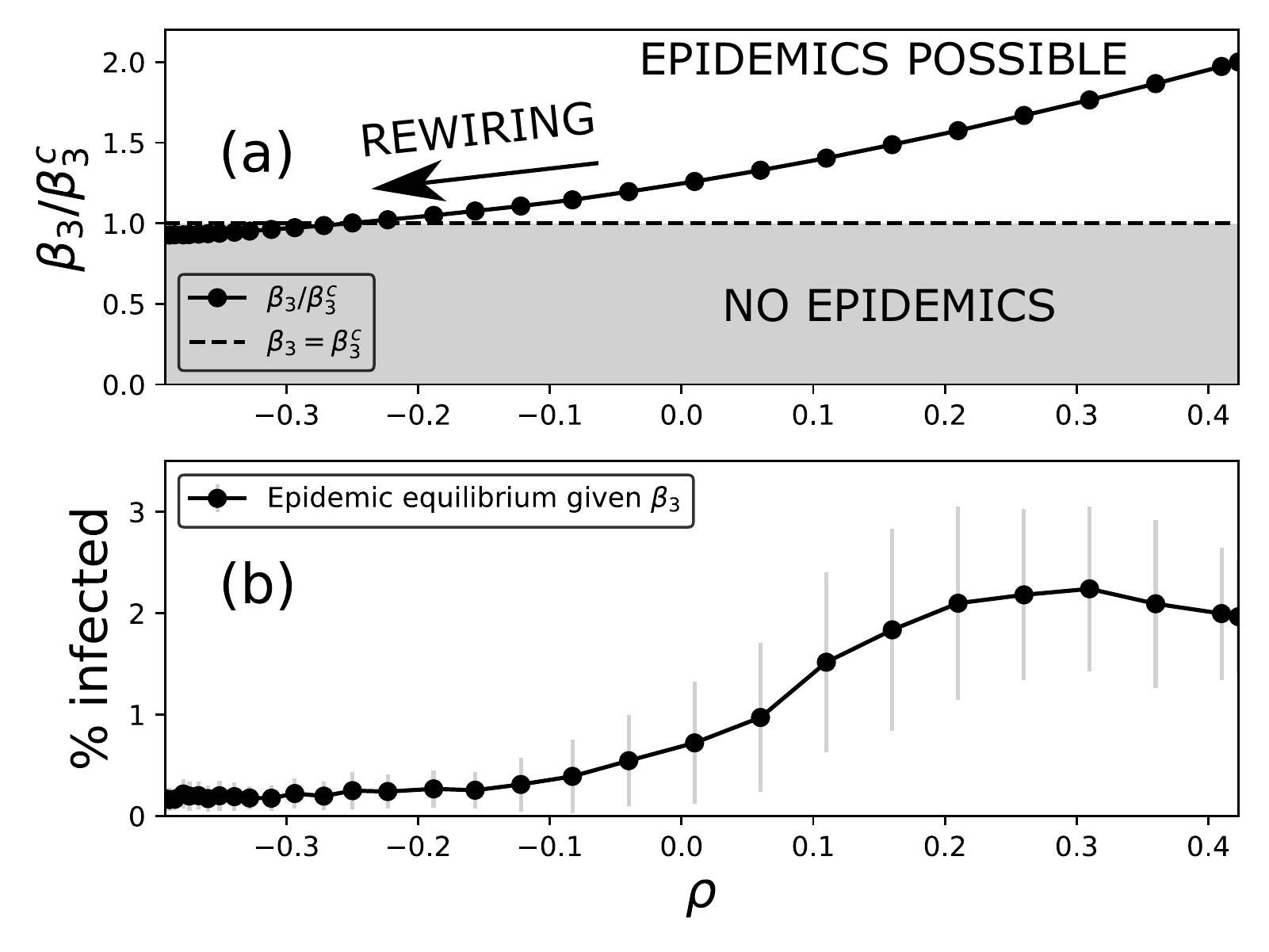}
    \caption{(a) The solid line with markers denotes the fixed value of $\beta_3$ as a fraction of the extinction threshold, $\beta_3/\beta_3^c$. The dashed line indicates $\beta_3/\beta_3^c=1$, below which epidemics are not possible. (b) The epidemic equilibrium (percentage of the population infected) for each hypergraph given the fixed value of $\beta_3$. The grey bars indicate the standard deviation at each data point.}
    \label{fig:epidemic_rewiring}
\end{figure}

We present an example based on the CB dataset, and additional cases in Appendix \ref{sec:appendix_rewiring_epidemics}. In this case, we consider $m=3$, $\gamma=1$, and $\beta_3 = 7.9 \times 10^{-3}$. In Fig.~\ref{fig:epidemic_rewiring}(a), we plot the chosen value of $\beta_3$ as a fraction of the extinction threshold, $\beta_3/\beta_3^c$ (solid line with markers), which decreases as $\beta_3^c$ is increased by hyperedge swaps, and the threshold for extinction (dashed line) $\beta_3/\beta_3^c=1$. Below the dashed line, epidemics are impossible. Above the dashed line, they may be possible. In Fig.~\ref{fig:epidemic_rewiring}(b), we plot the percentage of the population infected as a function of $\rho$ (averaged over 100 realizations of the epidemic). For more details about the numerical epidemic simulations see Appendix~\ref{sec:appendix_simulations}. For all values of $\rho$ such that $\beta_3/\beta_3^c<1$, no epidemics occur. For large enough values of $\rho$, however, we see that epidemics occur.

We caution, however, that decreasing $\lambda$ via hyperedge swaps might not necessarily suppress epidemics if $\beta_3/\beta_3^c$ is not reduced below 1. In principle, epidemics will occur for values of $\beta_3$ larger than a threshold $\beta_3^*\geq \beta_3^c$ which depends on the hypergraph structure. If the hyperedge swaps modify this threshold in such a way that $\beta_3^c < \beta_3^* <\beta_3$, when originally $\beta_3^c < \beta_3 < \beta_3^*$, epidemics can actually be promoted by the rewiring process (we show examples in Appendix \ref{sec:appendix_rewiring_epidemics}). Therefore, reduction of $\beta_3/\beta_3^c$ by preferential hyperedge swaps should be attempted only when one can guarantee that $\beta_3/\beta_3^c$ can be reduced below 1 or when there is already an epidemic.

\section{\label{sec:discussion} Discussion}

In this paper, we have presented a novel definition of assortativity for hypergraphs, related it to the expansion eigenvalue, and motivated its use in relating assortative structure in hypergraphs to the epidemic behavior. This approach, however, has limitations regarding the application of the expansion eigenvalue to hypergraphs and the calculation of the epidemic threshold.

There are two main limitations of the expansion eigenvalue. The first limitation is that one can think of the matrix associated to the right-hand side of Eq.~\eqref{eqn:expansion_eigenvalue} as the weighted adjacency matrix of an effective {\it pairwise} network, therefore reducing group interactions to multiple pairwise interactions. Such a reduction does not always capture all the complexity of nonlinear dynamical processes \cite{neuhauser_multibody_2020}. In particular, higher-order dynamical correlations might be missed by this approach. The second (related) limitation is that, since this eigenvalue is, by definition, a quantity related to linear processes, its applicability is restricted in principle only to certain dynamical regimes. However, approaches that reduce a hypergraph to an effective pairwise network have been successful and found application in clustering \cite{hayashi_hypergraph_2020}, diffusion and consensus \cite{jost_hypergraph_2019}, centrality \cite{benson_three_2019}, contagion \cite{bodo_sis_2016}, and other areas. In addition, as we showed, the expansion eigenvalue still encapsulates a large amount of information about the hypergraph structure, such as the hyperdegree distribution, correlations between degrees of different order, and assortative mixing. Therefore, the expansion eigenvalue should be considered as a complementary tool to other measures of hypergraph structure.

When deriving approximations to the epidemic threshold for the SIS model in pairwise networks, many approaches may be considered, such as using heterogeneous mean-field approaches \cite{boguna_epidemic_2002}, the largest eigenvalue of the adjacency matrix \cite{wang_epidemic_2003} (the quenched mean-field approach), the largest eigenvalue of the non-backtracking matrix \cite{karrer_percolation_2014}, the largest eigenvalue of the branching matrix \cite{karrer_percolation_2014}, message passing approaches \cite{karrer_message_2010,kirkley_belief_2021}, and many others. The differences between and advantages of these approaches are discussed at length in Ref.~\cite{wang_unification_2017}. The same is true for the epidemic threshold in hypergraphs. Although there are more accurate approximations of the epidemic threshold \cite{karrer_percolation_2014, goltsev_percolation_2008}, we employ the quenched mean-field approach because of its simple relation to the adjacency tensor of an $m$-uniform hypergraph and corresponding explainability. As in the pairwise network case, more sophisticated approaches \cite{matamalas_abrupt_2020} can yield a better approximation to the epidemic threshold.

Similarly, there are many ways to derive an approximation to the largest eigenvalue for pairwise networks, given an adjacency matrix. The authors in Ref.~\cite{chung_spectra_2003} derive the largest eigenvalue for a Chung-Lu network excluding any correlations. In Ref.~\cite{castellano_relating_2017}, the authors extend the approach of Ref.~\cite{chung_spectra_2003} by allowing degree-degree correlations to occur. We use a heterogeneous mean-field approach as in Ref.~\cite{chung_spectra_2003} which requires certain assumptions. First, we enforce that the hypergraph is realizable; that is, $f_m(k^{(m)}_1,\dots, k^{(m)}_m)\leq 1$ for every combination of $k^{(m)}_1,\dots, k^{(m)}_m$, which for the uncorrelated case requires that
\begin{equation*}
k^{(m)}_{max}\leq \sqrt[m]{\left(\sum_{i=1}^N k^{(m)}_i\right)^{m-1}}.
\end{equation*}
For the assortative case, this necessary condition depends on the specific assortativity function used. In addition, the mean-field approximation where we assume that nodes with the same hyperdegree have the same eigenvector entry is valid only when each node has a large number of connections, so that the states of neighbors of nodes with the same hyperdegree are statistically similar.

Despite these limitations, our results provide a way to connect various measures of hypergraph structure with dynamical processes in a systematic way and for a large class of tunable null models. We believe that exploring the role of the expansion eigenvalue in other dynamical processes on hypergraphs will be a fruitful research direction.

\section{\label{sec:code}Data Availability Statement} The data and code that support the findings of this study are openly available in \href{https://github.com/nwlandry/hypergraph-assortativity}{github.com/nwlandry/hypergraph-assortativity}~\cite{landry_2022_6415466}.

\vspace{0.1in}

\begin{acknowledgments}
Nicholas Landry would like to acknowledge a helpful conversation with Phil Chodrow.
\end{acknowledgments}

\appendix

\section{\label{sec:appendix_eigenvalue} More detailed derivation of the perturbed eigenvalue}

We start with the expansion of Eq.~(2) in the main text to first order (we recall that we are considering an $m$-uniform hypergraph), which is
\begin{equation}
\begin{aligned}
&\alpha \lambda^{(0)} k + \epsilon \lambda^{(0)} u_k^{(1)} + \alpha\epsilon\lambda^{(1)}k\\
&=\alpha \sum_{k_1,\dots,k_{m-1}} N(k_1)\dots N(k_{m-1})\\
&\times\frac{k\,k_1\dots k_{m-1}}{(N\langle k\rangle)^{m-1}}(k_1 + \dots + k_{m-1})\\
&+\epsilon \sum_{k_1,\dots,k_{m-1}} N(k_1)\dots N(k_{m-1})\\
&\times\frac{k\,k_1\dots k_{m-1}}{(N\langle k\rangle)^{m-1}}(u_{k_1}^{(1)} + \dots + u_{k_{m-1}}^{(1)})\\
&+\alpha\epsilon\sum_{k_1,\dots,k_{m-1}} N(k_1)\dots N(k_{m-1})\frac{k\,k_1\dots k_{m-1}}{(N\langle k\rangle)^{m-1}}\\
&\times g_m(k,k_1,\dots,k_{m-1})(k_1 + \dots + k_{m-1}).
\end{aligned}
\label{eq:first_order}
\end{equation}

From the 0th order approximation, the first terms on both sides of the equation are equal and we can cancel them. Secondly, assuming symmetry of $f_m$ and $g_m$, we can simplify the right-hand side as
\begin{align*}
&\epsilon \lambda^{(0)} u_k^{(1)} + \alpha\epsilon\lambda^{(1)}k\\
&=\epsilon (m-1)k\sum_{k_1}P(k_1)\frac{k_1 u_{k_1}^{(1)}}{\langle k\rangle}\\
&+\alpha\epsilon(m-1) \sum_{k_1,\dots,k_{m-1}} N(k_1)\dots N(k_{m-1})\\
&\times\frac{k\,k_1^2\, k_2\dots k_{m-1}}{(N\langle k\rangle)^{m-1}} g_m(k,k_1,\dots,k_{m-1}).
\end{align*}

We multiply both sides by $k \, P(k)/\langle k\rangle$ and sum over $k$ which yields
\begin{align*}
&\epsilon\lambda^{(0)} \sum_k P(k) \frac{k \, u_k^{(1)}}{\langle k\rangle} + \alpha\epsilon\lambda^{(1)}\sum_k P(k)\frac{k^2}{\langle k\rangle}\\
&= \epsilon(m-1)\sum_k P(k) \frac{k^2}{\langle k\rangle}\sum_{k_1}P(k_1)\frac{k_1 u_{k_1}^{(1)}}{\langle k\rangle}\\
&+\alpha \epsilon (m-1) \sum_{k,k_1,\dots,k_{m-1}} N(k)N(k_1)\dots N(k_{m-1})\\
&\times\frac{k^2\,k_1^2\,k_2\dots k_{m-1}}{(N\langle k\rangle)^{m}} g_m(k,k_1,\dots,k_{m-1}).
\end{align*}
Because $\lambda^{(0)}=(m-1)\langle k^2\rangle/\langle k\rangle$, the first terms on both sides are equal and we cancel them, yielding
\begin{align}
\epsilon\lambda^{(1)} &= \epsilon (m-1)\frac{\langle k\rangle}{\langle k^2\rangle}\sum_{k,k_1,\dots,k_{m-1}} N(k)N(k_1)\dots N(k_{m-1})\nonumber\\
&\times\frac{k^2\,k_1^2\dots k_{m-1}}{(N\langle k\rangle)^{m}} g_m(k,k_1,\dots,k_{m-1}).
\end{align}

We can use the relation that $f_m(k_1,\dots,k_m) = (m-1)!k_1\dots k_m/(N\langle k\rangle)^{m-1}\left[1 + \epsilon g_m(k_1, \dots, k_m)\right]$ to remove the reference to $g_m$, obtaining
\begin{align*}
\epsilon\lambda^{(1)} &= \frac{(m-1)}{(m-1)!}\frac{\langle k\rangle}{\langle k^2\rangle}\sum_{k,k_1,\dots,k_{m-1}} N(k)N(k_1)\dots N(k_{m-1})\\
&\times\frac{k\,k_1}{N\langle k\rangle} f_m(k,k_1,\dots,k_{m-1})\\
& - (m-1)\frac{\langle k\rangle}{\langle k^2\rangle}\sum_{k,k_1} P(k)P(k_1)\frac{k^2\,k_1^2}{\langle k\rangle^{2}}.
\end{align*}
The term 
\begin{align*}
&\frac{1}{2!(m-2)!}\sum_{k,k_1,\dots,k_{m-1}} N(k)N(k_1)\dots N(k_{m-1})\\
&\times k k_1 f_m(k,k_1,\dots,k_{m-1})
\end{align*}
represents the expected sum of all products of degrees for pairs of nodes belonging to the same hyperedge (where the factors $2!$ and $(m-2)!$ correct for overcounting permutations of $k,k_1$ and $k_2, k_3,\dots, k_{m-1}$ respectively). Since the number of possible pairwise products in an $m$-uniform hypergraph is given by
\begin{align}
\sum_{e\in E}\sum_{k,k'\in e, k\neq k'} 1 &=\binom{m}{2} |E| = \left( \frac{N\langle k\rangle}{m}\right)\left(\frac{m(m-1)}{2}\right)\nonumber\\
& = \frac{(m-1) N\langle k\rangle}{2},
\end{align}
letting $|E|$ be the number of edges, we can express $\lambda^{(1)}$ in terms of
\begin{align*}
\langle k k_1\rangle_E &= \frac{1}{(m-1)!}\sum_{k,k_1,\dots,k_{m-1}} N(k)N(k_1)\dots N(k_{m-1})\\
&\times\frac{k\,k_1}{N\langle k\rangle} f_m(k,k_1,\dots,k_{m-1}),
\end{align*}
the average of pairwise degree products over pairs of connected nodes, as
\begin{equation*}
\epsilon \lambda^{(1)} = (m-1)\frac{\langle k\rangle \langle k k_1\rangle_E}{\langle k^2\rangle} - \lambda^{(0)}.
\end{equation*}
Therefore,
\begin{equation}
\lambda = \lambda^{(0)} + \epsilon \lambda^{(1)} = (m-1)\frac{\langle k\rangle \langle k k_1\rangle_E}{\langle k^2\rangle} = \lambda^{(0)}(1 + \rho),
\end{equation}
where
\begin{equation}
\rho = \frac{\langle k\rangle^2 \langle k k_1\rangle_E}{\langle k^2\rangle^2} - 1.
\end{equation}

\section{\label{sec:appendix_rewiring_epidemics} Suppressing epidemics through preferential rewiring}

In this Section, we include additional plots of the effect of disassortative rewiring on the epidemic extent. We consider the CM and EE datasets described in the main text. The following plots have the same structure as that in the main text so we omit the legend for simplicity.

\begin{figure*}
\centering
\begin{tabular}{cc}
\subfloat[CM dataset, $\beta_3=1.78 \times 10^{-2}$ \label{fig:power-law}]{\includegraphics[width=8.6cm]{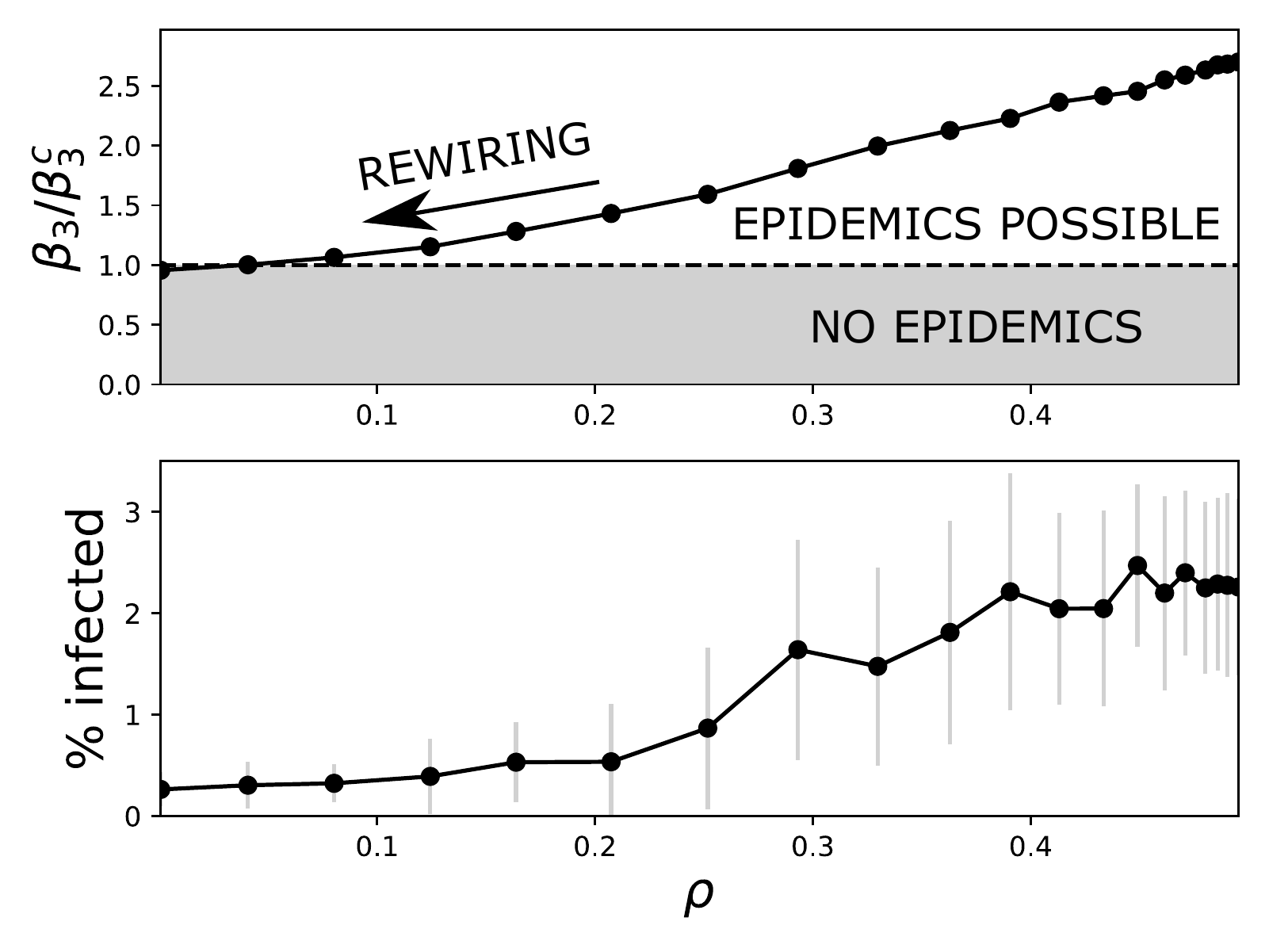}}
&
\subfloat[EE dataset, $\beta_3=1.3 \times 10^{-3}$ \label{fig:eu-emails_small_beta}]{\includegraphics[width=8.6cm]{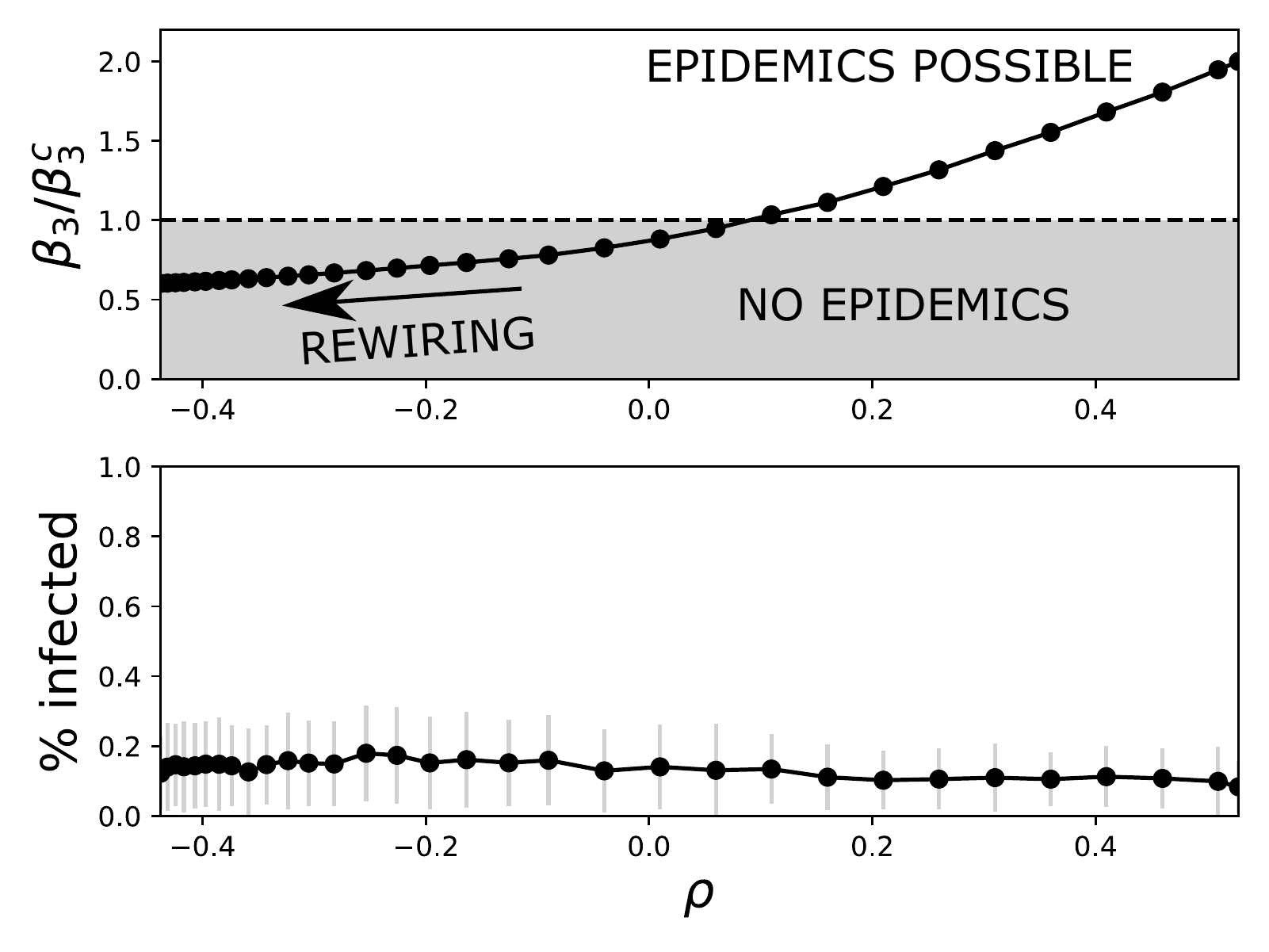}}
\\
\subfloat[EE dataset, $\beta_3=2.2 \times 10^{-3}$ \label{fig:eu-emails_medium_beta}]{\includegraphics[width=8.6cm]{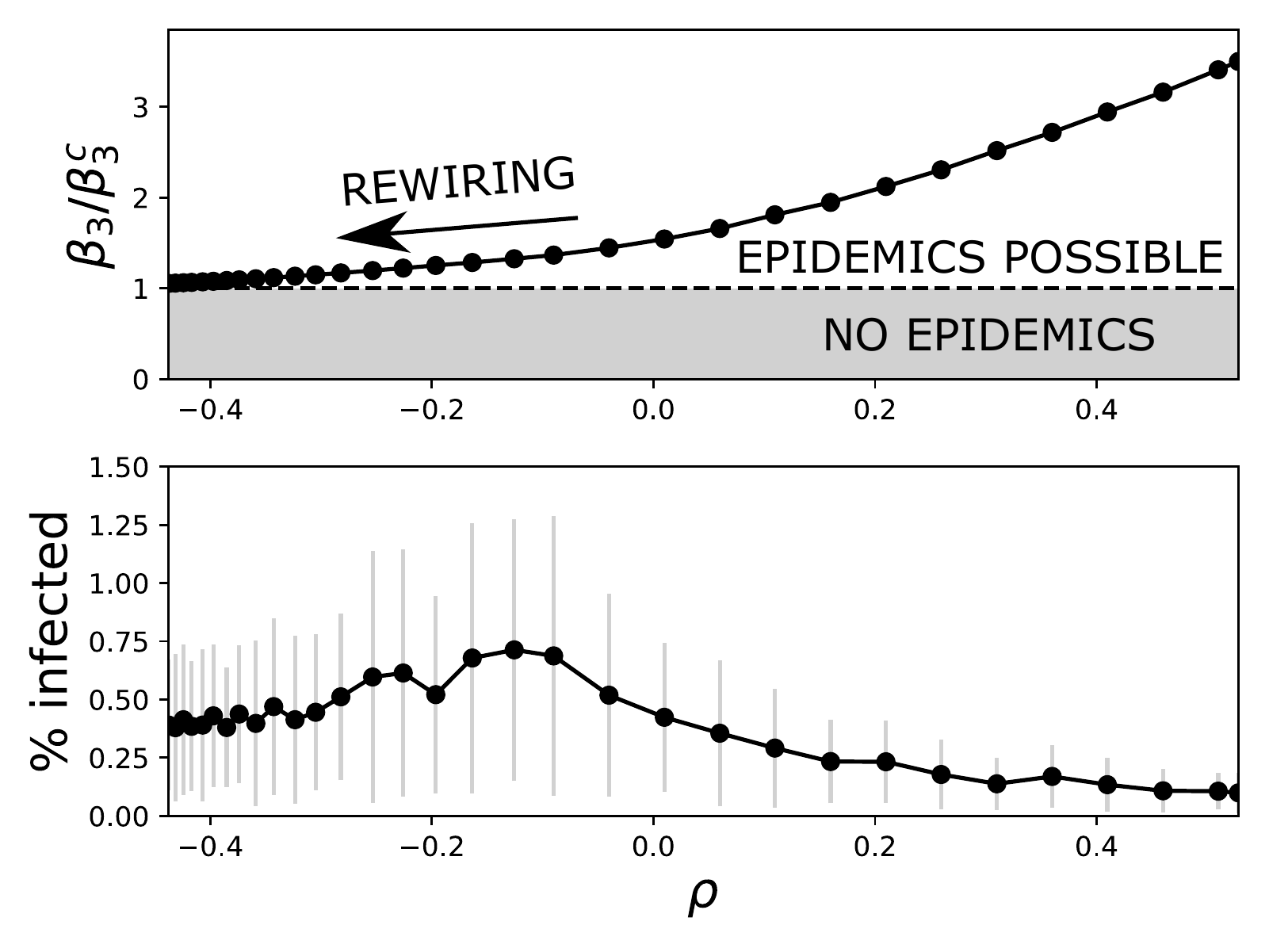}}
&
\subfloat[EE dataset, $\beta_3=3.2 \times 10^{-3}$ \label{fig:eu-emails_large_beta}]{\includegraphics[width=8.6cm]{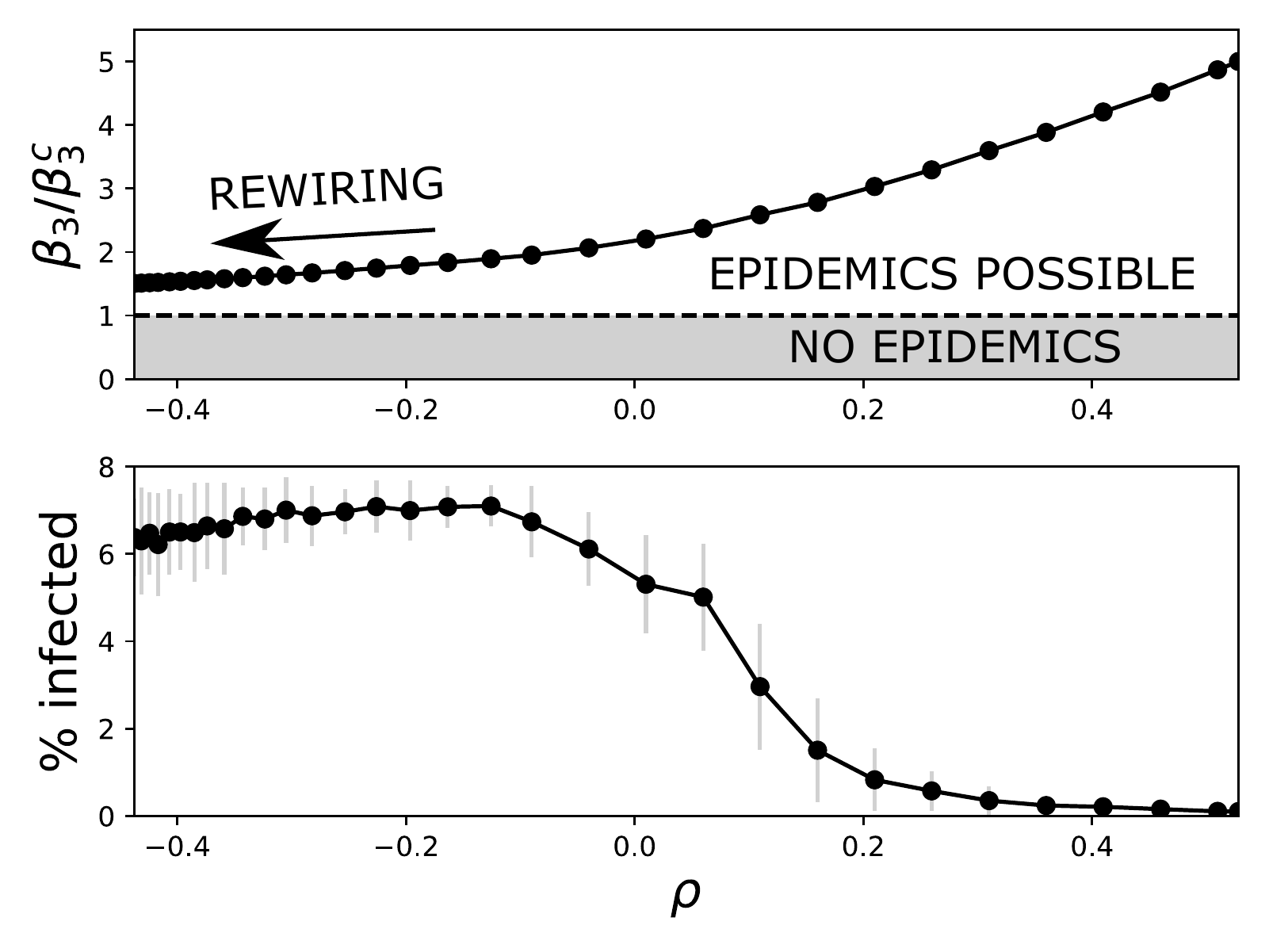}}
\end{tabular}
\caption{Additional plots of preferential rewiring and the corresponding epidemic response. Each subplot corresponds to a particular choice of infectious rate and dataset and follows the same format as Fig.~\ref{fig:epidemic_rewiring}.}
\label{fig:additional_rewiring}
\end{figure*}

In Fig.~\ref{fig:power-law}, we see the same behavior as that of the CB dataset. We comment that, as we expect, the epidemic threshold is fairly close to the predicted extinction threshold. In Figs.~\ref{fig:eu-emails_small_beta} and \ref{fig:eu-emails_large_beta}, we see behavior that differs from that of the CB dataset, but is consistent with our theoretical approach. In Fig.~\ref{fig:eu-emails_small_beta}, we see that the epidemic extent is roughly less than 0.25\% for all values of $\rho$. This does not contradict the bounds we derived because there is no epidemic below the extinction threshold. The behavior in Fig.~\ref{fig:eu-emails_large_beta} indicates that additional structure is present in the original hypergraph that seems to be suppressing the epidemic as well and warrants further study.

As discussed in the text, it is possible that if hyperedge swaps do not bring $\beta_3/\beta_3^c$ below 1 as in Fig.~\ref{fig:eu-emails_large_beta}, the process results in an epidemic. While we only see this for the EE dataset, one should be cautious of rewiring the hypergraph unless one can guarantee that $\beta_3/\beta_3^c < 1$ can be achieved.

\section{\label{sec:appendix_simulations} Numerical simulations} We model the hypergraph SIS contagion process as a continuous-time discrete-state (CTDS) Markov process, in contrast to Refs.~\cite{iacopini_simplicial_2019,landry_effect_2020} which assume a discrete-time (DTDS) process. In Ref.~\cite{burgio_network_2021}, the authors find that discrete-time processes inaccurately model contagion processes due to higher-order correlations. We note that as the time step in a DTDS process approaches zero, we recover the dynamics of the continuous time process.

As described in the main text, we consider a 3-uniform hypergraph of size $N$. We specify a spontaneous healing rate $\gamma$ and an infection rate $\beta_3$ at which an infected 3-hyperedge infects a susceptible node. The total rate at which infected nodes recover is given by the number of infected nodes $N_I$ multiplied by the healing rate $\gamma$. The rate at which each susceptible node $i$ is infected is given by the number of infected hyperedges (hyperedges with at least one infected neighbor) of which it is a member, $N^E_i$, multiplied by the infection rate $\beta_3$, and the total infection rate is $\beta_3\sum_{i=1}^N N^E_i$. The total rate at which these disjoint events occur is their sum, i.e., $R = \gamma N_I + \beta_3 \sum_{i=1}^N N^E_i$.

For this CTDS process, the time between events, $\tau$, is drawn from the exponential distribution with rate $R$. Once this time has been determined, we must determine which type of event occurred. The probability that a node recovers is $p = \gamma N_I/R$ and the probability that a node becomes infected is $1 - p$ and so we can draw the event from a Bernoulli distribution with parameter $p$. Next we must determine the node for which this event occurred. If an infected node has recovered, we select this node uniformly at random from the list of infected nodes. If a susceptible node has become infected, we select a node from the list of infected nodes according to the probabilities $p_i = N^E_i /\left(\sum_{i=1}^N N^E_i\right)$ for each node $i$.

Once the time increment, event type, and affected node have been determined, we first increment the time $t_i$ by $\tau$; second, we increment the number of infected individuals by one and decrease the number of susceptible individuals by one for an infection event (vice-versa for a recovery event); and lastly, we update the list of susceptible and infected nodes as well as the rates of each mechanism. We repeat these steps until either $t$ exceeds a maximum specified time or the number of infected nodes is zero. We refer to this termination time as $T$ and the corresponding number of discrete data points as $N_T$. Modeling the SIS contagion process as a CTDS process can be more efficient than a DTDS process when $R$ is small because the exponential distribution allows the simulation to take large steps in time when $R$ is small.

To recover the equilibrium from these simulations, we average over the last 10\% of the simulation time, i.e., the interval $[T_0, T]$, where $T_0=0.9T$. We calculate a weighted average of the number of infected nodes, where the weight on the first data point is proportional to the average interevent time in the interval $[T_0, T]$ and each subsequent weight is proportional to the interevent time between the previous data point and the current data point.

\section*{References}

\bibliography{hypergraph_assortativity}

\end{document}